\documentclass[prx,superscriptaddress,showkeys,twocolumn]{revtex4}
\usepackage{graphicx}
\usepackage{dcolumn}
\usepackage{rotating}
\usepackage{amssymb}
\usepackage{amsfonts}
\usepackage{amsmath}
\usepackage{bm}
\usepackage{mathptmx}

\begin{document}

\title {Motional Quantum States of Surface Electrons on Liquid Helium in a Tilted Magnetic Field}

\author{A. A. Zadorozhko}
\altaffiliation{These authors contributed equally to this work.}
\affiliation{Quantum Dynamics Unit, Okinawa Institute of Science and Technology (OIST) Graduate University, Onna, 904-0495 Okinawa, Japan}
\author{J. Chen}   
\altaffiliation{These authors contributed equally to this work.}
\affiliation{Quantum Dynamics Unit, Okinawa Institute of Science and Technology (OIST) Graduate University, Onna, 904-0495 Okinawa, Japan}
\author{A. D. Chepelianskii}
\affiliation{LPS, Univ. Paris-Sud, CNRS, UMR 8502, F-91405, Orsay, France}
\author{D. Konstantinov}
\email[Email: ]{denis@oist.jp}
\affiliation{Quantum Dynamics Unit, Okinawa Institute of Science and Technology (OIST) Graduate University, Onna, 904-0495 Okinawa, Japan}

\begin{abstract}

The Jaynes-Cummings model (JCM), one of the paradigms of quantum electrodynamics, was introduced to describe interaction between light and a fictitious two-level atom. Recently it was suggested that the JCM Hamiltonian can be invoked to describe the motional states of electrons trapped on the surface of liquid helium and subjected to a constant uniform magnetic field tilted with respect to the surface [Yunusova {\it et al}. Phys. Rev. Lett. \textbf{122}, 176802 (2019)]. In this case, the surface-bound (Rydberg) states of an electron are coupled to the electron cyclotron motion by the in-plane component of tilted field. Here we investigate, both theoretically and experimentally, the spectroscopic properties of surface electrons in a tilted magnetic field and demonstrate that such a system exhibits a variety of phenomena common to the light dressed states of atomic and molecular systems. This shows that electrons on helium realize a prototypical atomic system where interaction between components can be engineered and controlled by simple means and with high accuracy, and which therefore can be potentially used as a new flexible platform for quantum experiments. Our work introduces a pure condensed-matter system of electrons on helium into the context of atomic, molecular and optical physics.

\end{abstract}

\date{\today}


\maketitle

\section{Introduction}\label{intro}

The Jaynes-Cummings model (JCM) describes the interaction between a single two-level system and a quantum harmonic oscillator~\cite{Haro_book}. Originally formulated to account for the interaction between an atom and a single mode of the free-space electromagnetic radiation~\cite{JC}, this model became extensively used in quantum atomic, molecular and optical (AMO) physics. Recent developments in the experimental realization of this model, in particular within the setting of cavity quantum electrodynamics (CQED) which exploits the radiative coupling between a two-level atom and a cavity mode~\cite{Kimb2000,Remp2000,Schoe2004,Scho2004,Depp2004,Schu2019,Taka2020}, have opened a pathway for fundamental tests of quantum theory, such as non-destructive detection of quantum states,~\cite{Haro1999,Haro2007} realization of nonclassical states and observation of their decoherence~\cite{Mart2008,Haro2008}, and creation of many-body entanglement~\cite{Haro2000,Polz2001,Vule2010}. The main feature of CQED exploited in the experiments is a reversible Rabi oscillation between the coupled atom-photon states. These oscillations occur at a frequency determined by the rate of atom-photon coupling $g$ explicitly appearing in the JCM Hamiltonian. The 'strong coupling' regime of CQED is realized when the value of $g$ exceeds the decay rates for the atomic population and the cavity field. 

Under certain conditions, the JCM Hamiltonian of CQED can be formally applied to describe other systems. In particular, the harmonic motion of a trapped ion can represent a single mode of a cavity, while the ion's internal states can serve as a two-level atom. In this case, the coupling between internal and motional states is accomplished by the ion's motion through the spatially inhomogenious laser beams which are used to excite transitions between internal ionic states~\cite{Zoll1995,Wine2004}. The coherence of quantum states of laser-cooled ions can be preserved for many cycles of Rabi oscillations. Thus the strong coupling regime of CQED can be realized in experiments, which presents the cold ion system as another attractive platform for fundamental tests~\cite{Wine1996,Wine2000} and quantum information purposes~\cite{Blat2018}.

Free electrons trapped on the surface of liquid helium present a unique, extremely clean condensed-matter system which shares some striking similarities with atomic systems studied in AMO physics. The surface bound states of an electron on liquid helium are formed due to, on the one hand, an attraction to a weak image charge inside the liquid and, on the other hand, a repulsion from helium atoms, which prevents an electron to enter the liquid. So called Rydberg states of confined motion of such an electron perpendicular to the liquid surface have the energy spectrum similar to that of an electron in the hydrogen atom and can be spectroscopically studied by using microwave light~\cite{Grim1973,Coll2002}. At temperatures below 1~K, the dissipative decay rates for the excited Rydberg states are very low because they are limited only by the interaction of an electron with the capillary surface waves (ripplons)~\cite{Dykm2003,Yuri2007}. Surface electrons (SE) on liquid helium show interesting similarities with some well-known phenomena in Rydberg atoms,~\cite{Saff2009} such as the Coulomb shift of the transition frequency due to dipolar interaction between neighboring electrons~\cite{Lamb1980,Kons2009} and the Lamb shift of the Rydberg transition frequency due to interaction of SE with the quantum field of ripplons~\cite{Dykm2017,Coll2017}.   

The electron motion parallel to the surface of liquid is free. However, it can be confined to the quantized cyclotron orbits by applying a sufficiently strong magnetic field perpendicular to the surface~\cite{Grim1972}. Similar to an ion in a trap, the in-plane motion of an electron becomes harmonic, with an energy spectrum consisting of equidistant Landau levels tuned by the value of the applied magnetic field. Usually, the electron motions parallel and perpendicular to the surface are only weakly coupled via scattering of an electron from ripplons. However, a strong coupling can be induced by applying a magnetic field parallel to the surface~\cite{Grim1976}. Physically, the coupling is via the Lorentz force acting on an electron due to its in-plane motion and the parallel magnetic field. Recently, it was shown that the Hamiltonian of a surface electron in a magnetic field tilted with respect to the surface is formally equivalent to the Hamiltonian of an atom coupled to the quantum field of electromagnetic radiation, with the coupling constant $g$ proportional to the magnitude of the in-plane component of the applied magnetic field~\cite{Chep2019}. In such a case, the Rydberg orbital states serve as an atom, while the in-plane cyclotron motion represents a single mode of the electromagnetic field. The tunable mixing between the motional states of SE can be manifested by the shift in the transition frequency for the dressed Rydberg states, as indeed was observed in the experiment~\cite{Chep2019}. 

As was seen in the past, ability to engineer and control states of a quantum system weakly coupled to the environment can provide new platforms for fundamental studies and applications~\cite{Haro1999,Haro2007,Mart2008,Haro2008,Haro2000,Polz2001,Vule2010,Wine1996,Wine2000,Blat2018}. For this reason, electrons on helium in a tilted magnetic field can be a promising system to explore. In this paper, we theoretically and experimentally study the spectroscopic properties of such a system and show that it exhibits a variety of phenomena common to atomic and molecular systems interacting with a laser light, such as the eigenstate mixing, the light shift of energy levels, the Autler-Townes splitting, and the electromagnetically induced transparency. The predicted spectroscopic properties based on the JCM Hamiltonian contain no adjustable variables, while all the parameters which enter the theory can be readily controlled by simple means of constant electric and magnetic fields, which allows for a detail comparison with the experiment. We also show that the coherent dynamics of coupled motional states can dominate over the dissipative processes in the system. This presents electrons on helium as a new flexible platform for quantum experiments. 

This paper is organized as follows. Section~\ref{first} introduces the Hamiltonian of an electron on liquid helium in a tilted magnetic field and analyzes its eigenenergy spectrum. For the sake of clarity, the analysis is done analytically using appropriate approximations, as well as by numerical calculations. Section~\ref{second} presents an experiment where the spectroscopic properties of SE in tilted magnetic fields are studied by the Stark spectroscopy method. Also, a detailed comparison of the experimental results with the calculations is given. The discussion of our results and their implications for the future experiments are given in Section~\ref{third}.
 
\section{Background and model}\label{first}

Free electrons can be trapped near the surface of liquid helium due to, on the one hand, a weak attraction to the liquid due to polarizability of helium atoms and, on the other hand, a potential barrier at the vapor-liquid interface due to hard-core repulsion from the helium atoms arising from the Pauli exclusion principle. According to quantum mechanical principles, this allows such electrons to hover above the surface of liquid helium at a distance of about 10~nm, thus forming a two-dimensional (2D) electron system~\cite{Andrei_book,Monarkha_book}. The basic quantum-mechanical Hamiltonian for a single electron above liquid helium is given by

\begin{equation}
H = \frac{\textbf{P}^2}{2m_e} + V(\textbf{R}),
\label{sec1:H}
\end{equation}

\noindent where $m_e$ is the bare electron mass. Assuming an infinitely extended flat surface of liquid, the potential energy of an electron can be written as

\begin{equation}
V(\textbf{R}) = V_0 \Theta(-z) - \frac{\Lambda}{z} \Theta(z).
\label{sec1:V}
\end{equation}

\noindent Here $z$ is the electron coordinate in the direction perpendicular to the surface,  $V_0\sim 1$~eV is the height of the repulsive potential barrier at the vapor-liquid interface located at $z$=0, and $\Theta(z)$ is the Heaviside (step) function. The last term in \eqref{sec1:V} describes attraction of an electron to a weak image charge inside the liquid, where $\Lambda$ is determined by the dielectric constant of liquid helium $\epsilon$ (for vapor we assume $\epsilon=1$) as
 
\begin{equation}
\Lambda = \frac{e^2}{16\pi\epsilon_0} \left(\frac{\epsilon-1}{\epsilon+1}\right).
\end{equation}

\noindent Here, $\epsilon_0$ is the vacuum permittivity and $e>0$ is the elementary charge. The Hamiltonian \eqref{sec1:H} can be separated into two parts corresponding to the orbital motion of an electron in the direction perpendicular to the surface ($H_z$) and parallel to the surface. In $z$-direction, the electron motion is quantized into the surface bound states which are the eigenstates of the Hamiltonian 

\begin{equation}
H_z = \frac{p_z^2}{2m_e} + V_0 \Theta(-z) -\frac{\Lambda}{z} \Theta(z).
\label{sec1:Hz}
\end{equation}

\noindent The energy spectrum of this motion can be easily found by making a reasonable assumption of a rigid-wall repulsive barrier, that is $V_0\rightarrow +\infty$. In this case it coincides with the energy spectrum of an electron in the hydrogen atom $-R_e/n^2$, $n=1,2,..$ , where $R_e=m_e\Lambda^2/(2\hbar^2)$ is the effective Rydberg constant. This constant is about 63~meV (36~meV) for an electron above liquid $^4$He ($^3$He). An electron in the ground Rydberg state localizes above the surface of liquid at an average distance $\langle z\rangle \sim r_B$, where $r_B=\hbar^2/(\Lambda m_e)$ is the effective Bohr radius. This radius is about 7.8~nm (10.3~nm) for an electron above liquid $^4$He ($^3$He). 

In experiments, there is always a static electric field $E_\bot$ applied perpendicular to the liquid surface. Such a field serves as an effective positive charge background needed to neutralize the Coulomb repulsion between electrons. In addition, the dc Stark shift induced by $E_\bot$ provides a very convenient way to tune energy difference between the Rydberg states for their spectroscopic studies, as will be described in Section~\ref{second}. Such an applied field adds an additional term $eE_\perp z$ to the Hamiltonian \eqref{sec1:Hz} and changes its eigenenergy spectrum. For sufficiently small values of $E_\bot$, the energy shift for $n$-th Rydberg state is given by $eE_\bot z_{nn}$, where $z_{nn}$ is the mean value of the coordinate operator $z$ for this state. It is clear that the dc Stark shift is linear due to the inversion symmetry breaking in $z$ direction imposed by the repulsive barrier at the liquid surface. Already for moderate fields $E_\bot \sim 10$~V/cm the perturbation theory does not provide accurate estimates for the shifts, therefore one has to numerically solve the 1D eigenvalue problem with the Hamiltonian \eqref{sec1:Hz}.

The electron motion parallel to the surface is free with a continuous parabolic energy spectrum $p^2/(2m_e)$, where $\textbf{p}=p_x \textbf{e}_x + p_y \textbf{e}_y$ is the electron in-plane momentum and $\textbf{e}_i$, $i=x, y, z$ is the unit vector in $i$ direction. When SE are subject to a static magnetic field $\textbf{B}=B_z\textbf{e}_z$ applied perpendicular to the surface, the electron in-plane motion is quantized into the states with an equidistant energy spectrum $\hbar\omega_c(l+1/2)$ (the Landau levels), where $\omega_c=eB_z/m_e$ is the cyclotron frequency and $l=0,1,..$ is the quantum number. The vertical and in-plane motions of an electron are uncoupled, and the full Hamiltonian describing the electron's orbital motion can be represented as

\begin{equation}
H_0 =  H_z  +\frac{(\textbf{p} + e\textbf{A})^2}{2m_e} = H_z + \hbar\omega_c \left( a^\dagger a +\frac{1}{2} \right), 
\label{sec1:HperpB}
\end{equation}

\noindent where $\textbf{A}$ is the vector potential. Choosing the Landau gauge $\textbf{A}=B_zx\textbf{e}_y$, for which the eigenvalue of the momentum operator $p_y$ is a good quantum number, we define the operator $a=(\sqrt{2}l_B)^{-1}\left(p_xl_B^2/\hbar-i(x+x_0)\right)$, where $x_0=p_y/(eB_z)$ and $l_B=\sqrt{\hbar/{eB_z}}$. The operator $a$ satisfies the commutation relation $[a,a^\dagger]$=1. Each electron eigenstate is the product of a Rydberg state $|n\rangle$ of vertical motion corresponding to the eigenenergy $E_n$ of the Hamiltonian $H_z$ and a state $|l\rangle$ of in-plane cyclotron motion, where $a^\dagger a|l\rangle = l|l\rangle$. Throughout this paper we disregard the spin state of electron because the spin-orbit interaction for SE on liquid helium is negligibly small~\cite{Lyon2004}, so the spin degree of freedom is always uncoupled from the orbital motion. 

When an additional component of static magnetic field is applied parallel to the liquid surface, in other words when the magnetic field $\textbf{B}$ is tilted with respect to $z$ axis, the vertical and in-plane motions are coupled and the electron eigenstates are no longer the simple product states. For certainty, we consider the non-zero components of the field only in the $y$ and $z$ directions and use the Landau gauge for the corresponding vector potential $\textbf{A}=B_y z \textbf{e}_x + B_z x \textbf{e}_y$. The electron full Hamiltonian becomes

\begin{equation}
H=H_a+\hbar\omega_c \left( a^\dagger a +\frac{1}{2} \right)+\frac{\hbar\omega_y}{\sqrt{2}\l_B} \left( a^\dagger + a \right)z,
\label{sec1:HtiltB}
\end{equation}

\noindent where we define $\omega_y=eB_y/m_e$. The Hamiltonian for the orbital motion in $z$ direction includes now an additional (diamagnetic) term due to the parallel component of the magnetic field, that is

\begin{equation}
H_a=H_z+\frac{m_e\omega_y^2z^2}{2}=\sum\limits_n \varepsilon_\alpha |\alpha\rangle\langle \alpha|.
\label{sec1:Hz'}
\end{equation}

\noindent Here, $\alpha$ and $\varepsilon_\alpha$, $\alpha=1,2,..$ , are the Rydberg states and corresponding eigenenergies of the vertical motion renormilized due to the diamagnetic term~\cite{Chep2019}. In what follows, it will be more convenient to work in the product basis $|n,l\rangle$ of the Hamiltonian $H_0$ given Eq.~\eqref{sec1:HperpB}, with corresponding energy eigenvalues $E_n+\hbar\omega_c(l+1/2)$, rather than using the renormalized product basis $|\alpha,l\rangle$.

The last term in the Hamiltonian \eqref{sec1:HtiltB} presents the coupling (paramagnetic) term due to the parallel component of the magnetic field. It arises because of the interaction between the magnetic dipole moment of an electron due to the orbital angular momentum $p_xz$ and the magnetic field $B_y$. The Hamiltonian \eqref{sec1:HtiltB} is reminiscent of that of an atom which interacts with a quantum field $a^\dagger a$ of an electromagnetic mode via the electrical dipole moment $ez$ of the atom~\cite{Cohen_book}. Thus, our system can be thought of as an atom with the Rydberg states of the renormalized Hamiltonian \eqref{sec1:Hz'} coupled to the bosonic field of the electron cyclotron motion, and with the coupling rate $g$ tuned by the value of the in-plane field $B_y$~\cite{Chep2019}. However, we note that the inversion symmetry breaking for $z$ direction in the SE system makes the situation somewhat different. In particular, the non-zero diagonal matrix elements $z_{nn}$ make contributions to the eigenvalues and eigenstates of the coupled system described by Eq.~\eqref{sec1:HtiltB}, while it is usually not the case for an atom-in-cavity system.

\begin{figure}
\includegraphics[width=8.6cm]{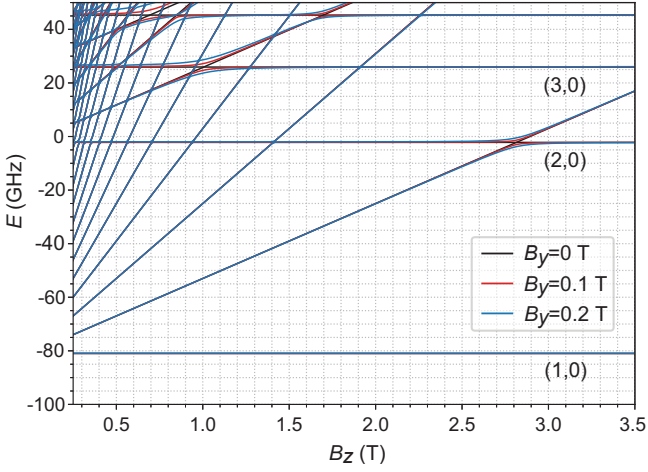}
\caption{(color online) Energy eigenvalues (in GHz) of the Hamiltonian given by Eq.~\eqref{sec1:HtiltB} (minus $\hbar\omega_c/2$) for an electron on liquid $^3$He in a perpendicular electric field $E_\bot=15$~V/cm versus the magnetic field $B_z$. Different colors correspond to the different values of the coupling field $B_y=0$ (black lines), 0.1 (red lines), 0.2~T (blue lines). The lowest ($l=0$) energy levels for each $n$-th manifold are indicated by the collective indexes $(n,0)$.}
\label{fig:2}       
\end{figure}

\begin{figure}
\includegraphics[width=8.6cm]{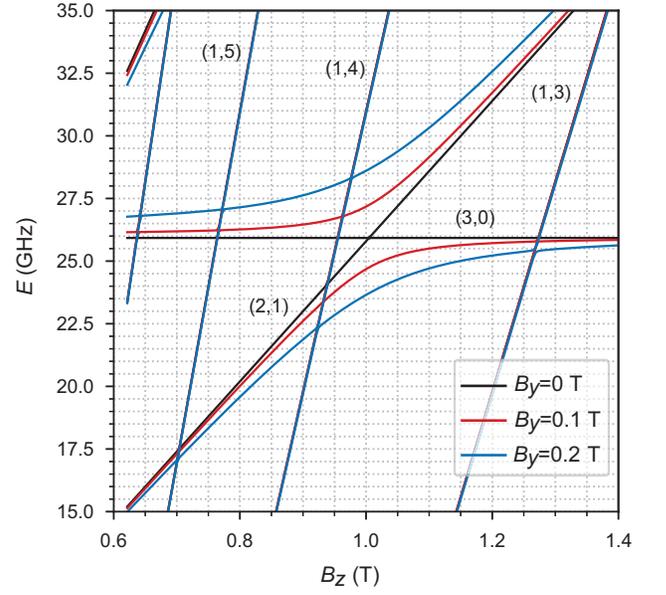}
\caption{(color online) Energy eigenvalues (in GHz) of the Hamiltonian given by Eq.~\eqref{sec1:HtiltB} (minus $\hbar\omega_c/2$) versus the magnetic field $B_z$ illustrating the avoided crossing of energy levels $(2,1)$ and $(3,0)$ which happens at non-zero coupling fields $B_y$. Several energy levels of the $n=1$ manifold are also indicated by the collective indexes $(1,l)$. The plots are obtained under the same conditions as for Fig.~\ref{fig:2}.}
\label{fig:3}       
\end{figure}

The eigenvalues and eigenstates of the Hamiltonian \eqref{sec1:HtiltB} can be obtained numerically, for example by the diagonalization of the matrix representation of $H$ constructed on a sufficiently large Hilbert sub-space. As mentioned earlier, we prefer to use the product basis $|n,l\rangle$ of the eigenstates of the Hamiltonian \eqref{sec1:HperpB}. Also, for the sake of comparison with experiment (Section~\ref{second}), where the energy spectrum of SE is probed by looking at the microwave-excited transitions between the lowest energy level and the higher energy levels, it is convenient to subtract the ground-state energy of the cyclotron motion $\hbar\omega_c/2$ from the full Hamiltonian \eqref{sec1:HtiltB}. Figure~\ref{fig:2} shows the energy eigenvalues of the Hamiltonian versus the perpendicular magnetic field $B_z$ for several values of $B_y$. The eigenvalues are obtained by the numerical diagonalization of the Hamiltonian matrix constructed in a subset of $|n,l\rangle$ with $1\leq n \leq 6$ and $0\leq l \leq 50$. For $B_y=0$, that is when there is no coupling between the vertical and in-plane motion of SE, there is a manifold of energy levels $E_{n,l}=E_n + \hbar\omega_c l$ for each $n$-th Rydberg state, each shifting linearly with $B_z$ and having a slope proportional to $l$. For the sake of clarity, the lowest energy levels ($l$=0) for each manifold are marked in Fig.~\ref{fig:2} by the collective indexes $(n,0)$. The largest effect of the non-zero $B_y$ can be seen at the crossings of energy levels of different manifolds. In particular, the coupling leads to the avoided crossing of energy levels, which implies the mixing between the corresponding product states. Fig.~\ref{fig:3} shows a magnified fragment of Fig.~\ref{fig:2} illustrating such an avoided crossing between energy levels $(3,0)$ and $(2,1)$ corresponding to the first ($l=0$) Landau level of the third ($n=3$) Rydberg state manifold and the second ($l=1$) Landau level of the second ($n=2$) Rydberg state manifold, respectively. In addition, there are significantly weaker anti-crossings between these levels and the energy levels of the $n=1$ manifold indicated by $(1,l)$ for several values of $l$.

Following our analogy with an atom in a cavity, we can talk about 'dressed' states of the electron orbital motion whose energies at the crossing point are separated by a gap proportional to the coupling strength. The dressed states are mixtures of the product states $|n,l\rangle$, thus in general are entangled states of the electron orbital motion perpendicular and parallel to the liquid surface. Of particular interest is the crossing between two levels for which the quantum numbers $l$ and $l'$ differ by 1. The region of $B_z$ near the crossing of such energy levels corresponds to the resonant regime of coupling in CQED. In this regime, we can obtain essential results in an analytical form by considering only a subspace of two nearly degenerate eigenstates $|n,l+1\rangle$ and $|n',l\rangle$ of the uncoupled system and perform Hamiltonian diagonalization in this subspace. The treatment is similar to the two-level atom coupled to a quantum electromagnetic mode, that is JCM~\cite{Scho2004}. It is convenient to introduce the coupling constant $g_{nn'}$ defined by the coupling matrix element of the interaction Hamiltonian $H_I$ given by he last term in \eqref{sec1:HtiltB}

\begin{equation}
\langle n,l| H_I |n',l'\rangle = g_{nn'} \sqrt{l+1} \delta_{l+1,l'} + g_{nn'} \sqrt{l} \delta_{l-1,l'}, 
\label{sec1:matrix}
\end{equation}    

\noindent where $\delta_{l,l'}$ is the Kronecker delta. Thus, we obtain

\begin{equation}
g_{nn'}=\frac{\hbar\omega_y}{\sqrt{2}}\frac{z_{nn'}}{l_B}=\sqrt{\frac{\hbar m_e\omega_c\omega_y^2}{2}}z_{nn'}.
\label{sec1:g}
\end{equation}
 
\noindent Under the above assumptions, the eigenstates of the Hamiltonian are given by

\begin{subequations}
\begin{align}
& |+,l\rangle = \cos (\theta_l/2) |n',l\rangle + \sin (\theta_l/2) |n,l+1 \rangle, \\
& |-,l\rangle = -\sin (\theta_l/2) |n',l\rangle + \cos (\theta_l/2) |n,l+1 \rangle,
\end{align} \label{sec1:dressed} 
\end{subequations}

\noindent where the 'mixing angle' $\theta_l$ is given by 

\begin{equation}
\theta_l= \tan^{-1} \left( \frac{g_{nn'}\sqrt{l+1}}{2E_\Delta} \right),
\label{sec1:angle}
\end{equation}  

\noindent and the corresponding energy eigenstates are given by

\begin{equation}
E_{\pm} = E_\Sigma \pm \sqrt{E_\Delta^2 + (l+1)|g_{nn'}|^2}.
\label{sec1:eigen}
\end{equation}   

\noindent Here, $E_{\Sigma(\Delta)}=\left( \tilde{E}_{n,l+1}\pm \tilde{E}_{n',l}\right)/2$, $\tilde{E}_{n,l}=E_{n,l}+m_e\omega_y(z^2)_{nn}/2$, and $(z^2)_{nn}$ is the mean value of $z^2$ for the $n$-th Rydberg state. The energy splitting between the dressed states \eqref{sec1:dressed} at the energy crossing for uncoupled states ($E_\Delta=0$) is given by $2\sqrt{l+1}|g_{nn'}|$. Note that the scaling of this splitting with $l$ is similar to the scaling of the Rabi splitting in CQED, where it changes with the number of photons in the cavity $n_\textrm{ph}$ as $\sqrt{n_\textrm{ph}}$~\cite{Scho2004}. In addition, the splitting increases linearly with the coupling field $B_y$ and the transition dipole moment $z_{nn'}$.

As can be seen in Fig.~\ref{fig:3}, for the crossings between two levels for which the quantum numbers $l$ and $l'$ differ by more than 1, there are significantly weaker anti-crossings.  This effect comes from the higher-order mixing between different eigenstates, which can be captured by the diagonalization of the Hamiltonian \eqref{sec1:HtiltB} on a sufficiently large sub-space of the product basis $|n,l\rangle$.
        
\begin{figure}
\includegraphics[width=7.0cm]{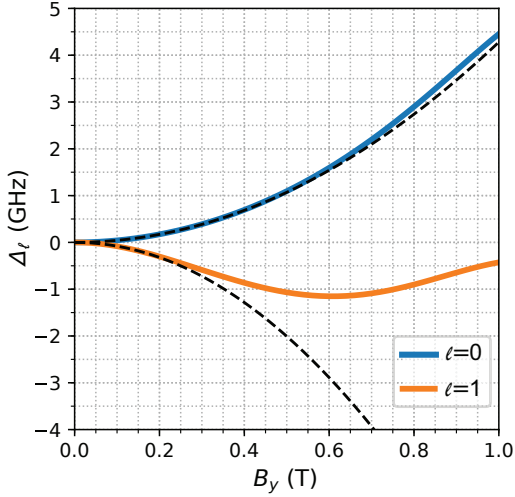}
\caption{(color online) The Lamb shift ($\Delta_0$) and light shift ($\Delta_1$) of the Rydberg $n=1\rightarrow 2$ transition frequency versus the coupling field $B_y$ obtained numerically (solid lines) and using the perturbation theory (dashed lines) for an electron on liquid $^3$He in the perpendicular electric field $E_\bot=15$~V/cm and perpendicular magnetic field $B_z=0.65$~T.} 
\label{fig:4}       
\end{figure} 

Even far from the level crossing, which corresponds to the non-resonant (dispersive) regime of coupling in CQED~\cite{Scho2004}, there is an appreciable shift of the energy levels induced by the state mixing. This is analogous to the 'light shift' of the atomic transition frequency experienced by an atom in the detuned cavity field~\cite{Haro_book}. It is instructive to consider the case where the interaction term $H_I$ in \eqref{sec1:HtiltB} can be treated as a perturbation. Also, we will assume sufficiently small values of $B_y$ such that the diamagnetic term $m_e\omega_y^2z^2/2$ can be treated as a perturbation as well. As before, it is convenient to choose the eigenstates $|n,l\rangle$ of the Hamiltonian \eqref{sec1:HperpB} as an unperturbed basis, therefore treat $H_1=H_I+m_e\omega_y^2z^2/2$ as the perturbation. To the lowest order in $B_y$, the shift of the unperturbed energy level $E_{n,l}$ is the sum of the first-order correction $\delta E_{n,l}^{(1)}$ due to the diamagnetic term and the second-order correction $\delta E_{n,l}^{(2)}$ due to the interaction term

\begin{equation}
\Delta E_{n,l} = \frac{m_e\omega_y^2 (z^2)_{nn}}{2} + \frac{\hbar^2\omega_y^2}{2l_B^2} \sum\limits_{n',l'} \frac{|z_{nn'}|^2|\langle l'| a^\dagger+a|l\rangle |^2}{E_{nn'}+\hbar\omega_c(l-l')},
\label{sec1:Eshift}
\end{equation}
      
\noindent where $E_{nn'}=E_n-E_{n'}$. Interestingly, there is a strong cancellation between the first and the second terms in \eqref{sec1:Eshift}, which is reminiscent of the calculations of the Lamb shift in the Hydrogen atom~\cite{Beth1947} and the ripplonic Lamb shift in electrons on helium~\cite{Dykm2017}. This can be seen by using the Bethe-type approach, that is expanding the mean value $(z^2)_{nn}$ in the first term in \eqref{sec1:Eshift} using the completeness relation for the eigenstates $|n\rangle$, that is $(z^2)_{nn}=\sum_{n'}|z_{nn'}|^2$. Plugging this expansion in \eqref{sec1:Eshift}, it is clear that the leading term proportional to $|z_{nn}|^2$ cancels out. The remaining shift reads

\begin{equation}
\Delta E_{n,l} = \frac{m_e\omega_y^2}{2} \sum\limits_{n'\neq n} |z_{nn'}|^2 \left( 1 + \frac{\hbar\omega_cl}{E_{nn'}+\hbar\omega_c} + \frac{\hbar\omega_c(l+1)}{E_{nn'}-\hbar\omega_c} \right).
\label{sec1:dE}
\end{equation}

As a particular example, let us consider the frequency shift for the transition between the ground ($n=1$) and the first-excited ($n=2$) Rydberg states. For an electron occupying the $l$-th state of the cyclotron motion, the corresponding shift in the transition frequency is given by $\Delta_l = (\Delta E_{2,l}-\Delta E_{1,l})/h$. In particular, for $l=0$ we obtain

\begin{equation}
\Delta_0 = \frac{m_e\omega_y^2}{2h} \left( \sum\limits_{n\neq 2} |z_{2n}|^2 \frac{E_{n2}}{E_{n2}+\hbar\omega_c} - \sum\limits_{n\neq 1} |z_{1n}|^2 \frac{E_{n1}}{E_{n1}+\hbar\omega_c} \right).
\label{sec1:D0}
\end{equation}         

\noindent This shift is an analogue of the Lamb shift in an atom due to its interaction with the vacuum ($n_\textrm{ph}=0$) cavity field~\cite{Haro_book}. In the experiment, the Lamb shift can be observed by exciting the Rydberg $n=1\rightarrow 2$ transition in SE occupying the lowest ($l=0$) Landau level (Section~\ref{second}). Similar expressions can be obtained for the light shifts $\Delta_l$ for SE occupying the higher Landau levels. Figure~\ref{fig:4} shows the corresponding frequency shifts $\Delta_0$ and $\Delta_1$ (in GHz) obtained using the perturbation approach as described above (dashed lines). To illustrate validity of this perturbation approach, the corresponding shifts obtained from the diagonalization of the full Hamiltonian \eqref{sec1:HtiltB} are plotted by the solid lines. We note that both shifts were experimentally confirmed earlier~\cite{Chep2019}. 

To summarize this section, the calculated eigenstates and energy eigenvalues of SE in a tilted magnetic field allows us to make certain predictions regarding the spectroscopic properties of this system. In particular, we predict certain features arising from the mixing of the electron motional states, such as the off-resonance shifts and resonant avoided crossings in the energy spectrum of SE, which can be probed in an experiment which employs spectroscopic methods. 

\section{Experimental setup and results}\label{second}

\begin{figure}
\includegraphics[width=8.6cm]{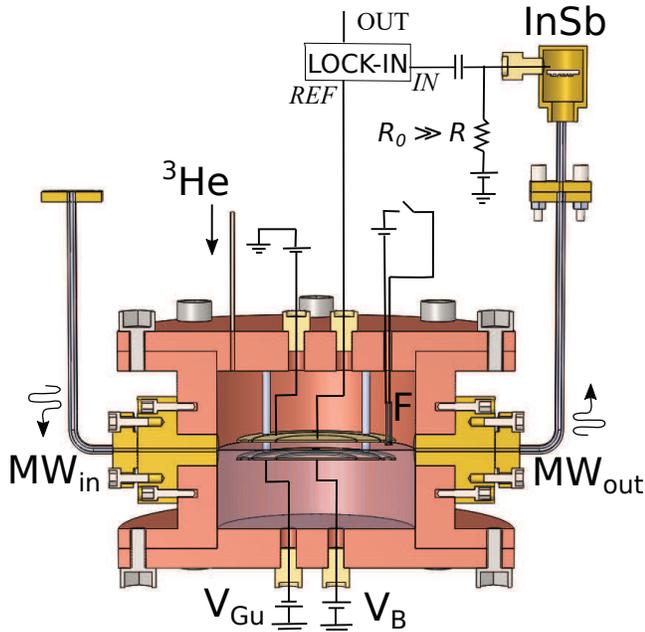}
\caption{(color online) Schematic 3D view of the experimental cell. Details are provided in the text.} 
\label{fig:0}       
\end{figure}

To check predictions of our calculations regarding the energy spectrum of SE, we performed an experiment with electrons on liquid $^3$He using the Stark spectroscopy method employed earlier~\cite{Grim1973,Kons2013,Coll2017}. The experiment is performed in a leak-tight cylindrical copper cell (see Fig.~\ref{fig:0}) cooled down to temperatures below 1~K in a dilution refrigerator. The cell is placed inside a superconducting vector magnet (not shown) which can produce a static magnetic field in both $y$ (horizontal) and $z$ (vertical) directions. The cell can be filled with the $^3$He gas through a thin capillary tube from a room temperature storage tank. The cell has two side windows located opposite to each other and fitted with home-made microwave (MW) waveguide flanges. The waveguide flanges serve as input and output ports for MW radiation which is used to excite transitions between energy eigenstates of SE. Both windows are sealed with the MW-transparent Kapton film using the Stycast epoxy to prevent leakage of helium from the cell into the vacuum space of the refrigerator.
    
Inside the cell, there are two round metal discs of diameter $D=30$~mm which form a parallel-plate capacitor with the distance between the disks $d=2$~mm. This distance is determined by the height of four cylindrical quartz spacers (not shown in Fig.~\ref{fig:0}) placed between the disks. In addition, each disk is divided into three concentric electrodes by two circular gaps of diameters 18 and 24~mm, each having a width of about 0.2~mm. $^3$He gas, which is introduced into the cell, is condensed in the cooled cell until the surface of liquid helium covers the bottom disk and the liquid level is set approximately in the middle between the bottom and top disks. Free electrons are injected into the space above the surface of liquid from a tungsten filament F by thermionic emission, while applying a positive voltage $V_\mathrm{B}$ to the inner circular electrode of the bottom disk. As a result, the injected electrons are attracted towards the liquid surface and form a round pool with typical areal density $n_s$ of the order $10^7$~cm$^{-2}$ on the surface just above the positively biased bottom electrode. With a positive voltage applied, SE can be held on the surface of liquid helium indefinitely long. In addition, the middle electrodes of the top and bottom disks serve as guard rings. By applying a negative voltage $V_\mathrm{Gu}$ to the guard rings, electrons can be stronger confined on the liquid surface to prevent their escape to the grounded walls of the experimental cell. 

To excite transitions between the Rydberg states of SE, the MW radiation in the 100~GHz range is transmitted into the cell from a room-temperature source (not shown in Fig.~\ref{fig:0}) through a waveguide coupled to the input port of the cell. In the experiment, the frequency $\omega$ of the radiation is fixed, while the transition frequency of SE is tuned to match the MW frequency by sweeping the perpendicular electric field $E_\bot = V_\mathrm{B}/d$ (dc Stark effect). The MW absorption due to resonant transitions induced in SE is measured as a change in the power of radiation transmitted through the cell. In order to measure the transmitted power, the output MW port of the cell is coupled to a cryogenic InSb bolometer (QMC Instruments Ltd.) operating at the temperature of the mixing chamber of the dilution refrigerator. The bolometer changes its resistance $R$ when it is heated by the incident radiation. To observe change in the bolometer resistance, therefore the incident MW power, we pass a dc current $I\approx 100$~$\mu$A generated by a battery and measure the voltage drop $IR$ across the bolometer. To increase sensitivity of the method, we apply a small modulating ac voltage $V_\mathrm{ac}=40$~mV$_\mathrm{rms}$ at the frequency $f_\mathrm{m}\approx 10$~kHz to the central electrode of the top disk of the parallel-plate capacitor. Due to the Stark shift, this modulates the detuning of the transition frequency of SE with respect to the MW frequency, therefore the MW power absorbed by SE and the MW power transmitted through the cell. The corresponding modulation of the voltage across the bolometer is then detected using the conventional lock-in amplifier operated at the modulation frequency $f_\mathrm{m}$.

\begin{figure}
\includegraphics[width=8.0cm]{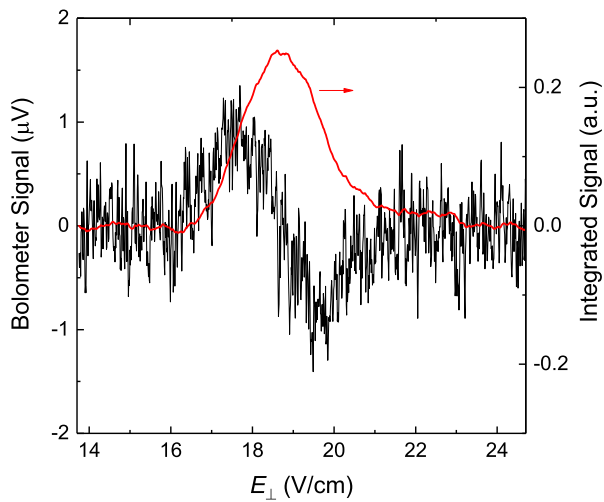}
\caption{(color online) Bolometer signal recorded at $T=0.33$~K for electron density $n_s=5\times 10^6$~cm$^{-2}$ by sweeping the perpendicular electric field $E_\bot$ to tune the $n=1\rightarrow 2$ transition between the Rydberg states of SE at $B_z=0.7$~T and $B_y=1$~T in resonance with MWs at frequency $\omega/2\pi=90$~GHz. The red line is an integrated signal which gives the inhomogeneously broadened absorption line for the $n=1\rightarrow 2$ transition.} 
\label{fig:5}       
\end{figure}

An example of the bolometer signal recorded by the lock-in amplifier for SE at $B_z=0.7$~T, $B_y=1$~T, and under presence of MW radiation at the frequency $\omega/2\pi=90$~GHz is shown in Fig.~\ref{fig:5}. For this frequency, the observed signal corresponds to the transition from the ground ($n=1$) state to the first excited ($n=2$) Rydberg state of SE. By keeping the amplitude of the modulating voltage $V_\mathrm{ac}$ to be much smaller than the width of the transition line, we record the derivative of the absorption line. Then, the absorption line can be obtained from the recorded bolometer signal by numerical integration (the red line in Fig.~\ref{fig:5}). All data presented here were taken at cell temperatures $T= 0.3$-$0.4$~K, measured by a calibrated ruthenium-oxide chip attached to the cell's top. The width of the line is mostly determined by the inhomogeneous broadening due to non-uniformity of $E_\bot$, which is much lager than the intrinsic linewidth of the transition line due to the scattering from ripplons~\cite{Issh2007}. Note that the power of MW excitation was kept very low in the experiment described here in order to avoid strong heating of SE and heating-related effects associated with electron-electron interaction~\cite{Kons2009}. 

\begin{figure}[h]
	\centering
	\includegraphics[width=1.1\linewidth]{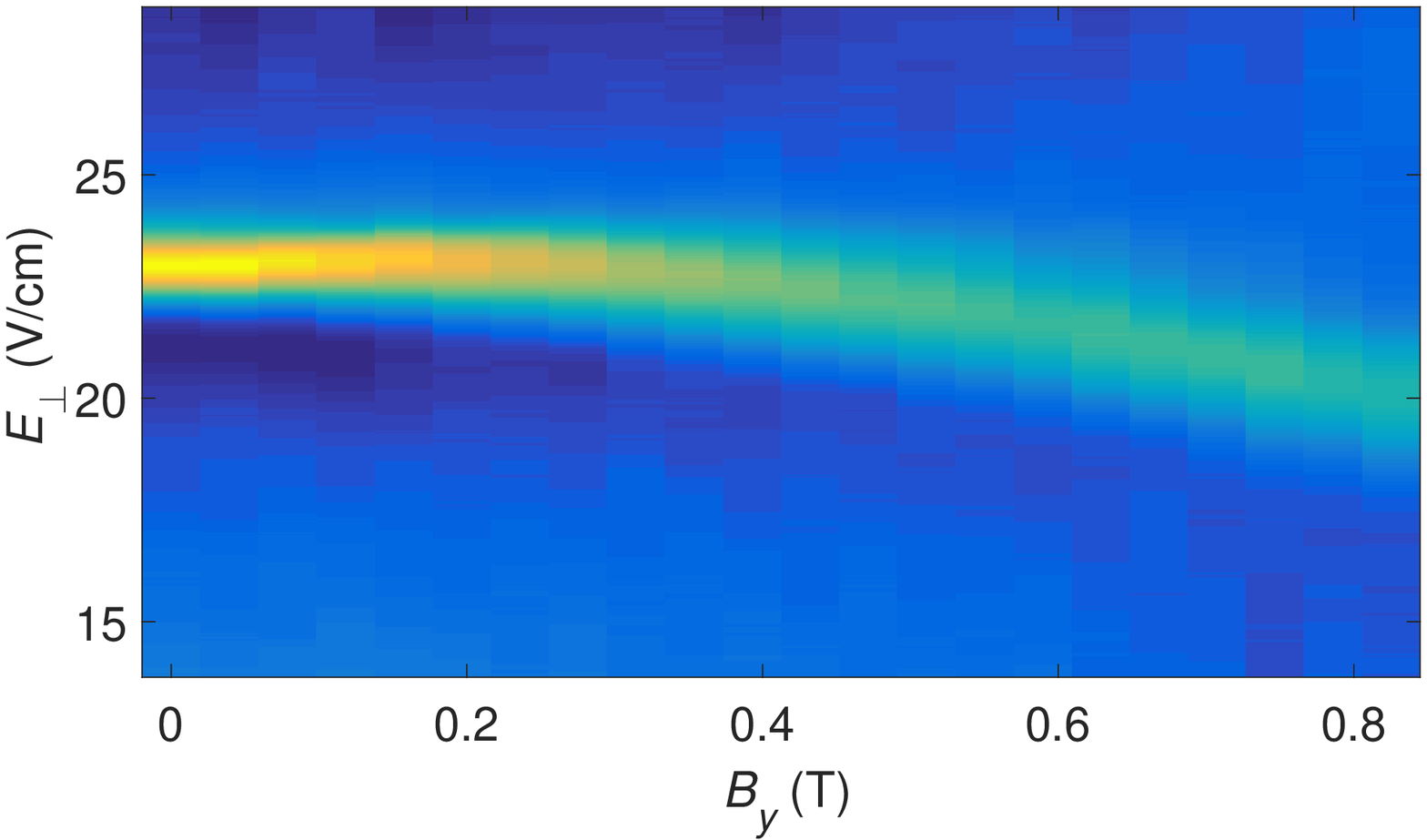}
	\includegraphics[width=1.1\linewidth]{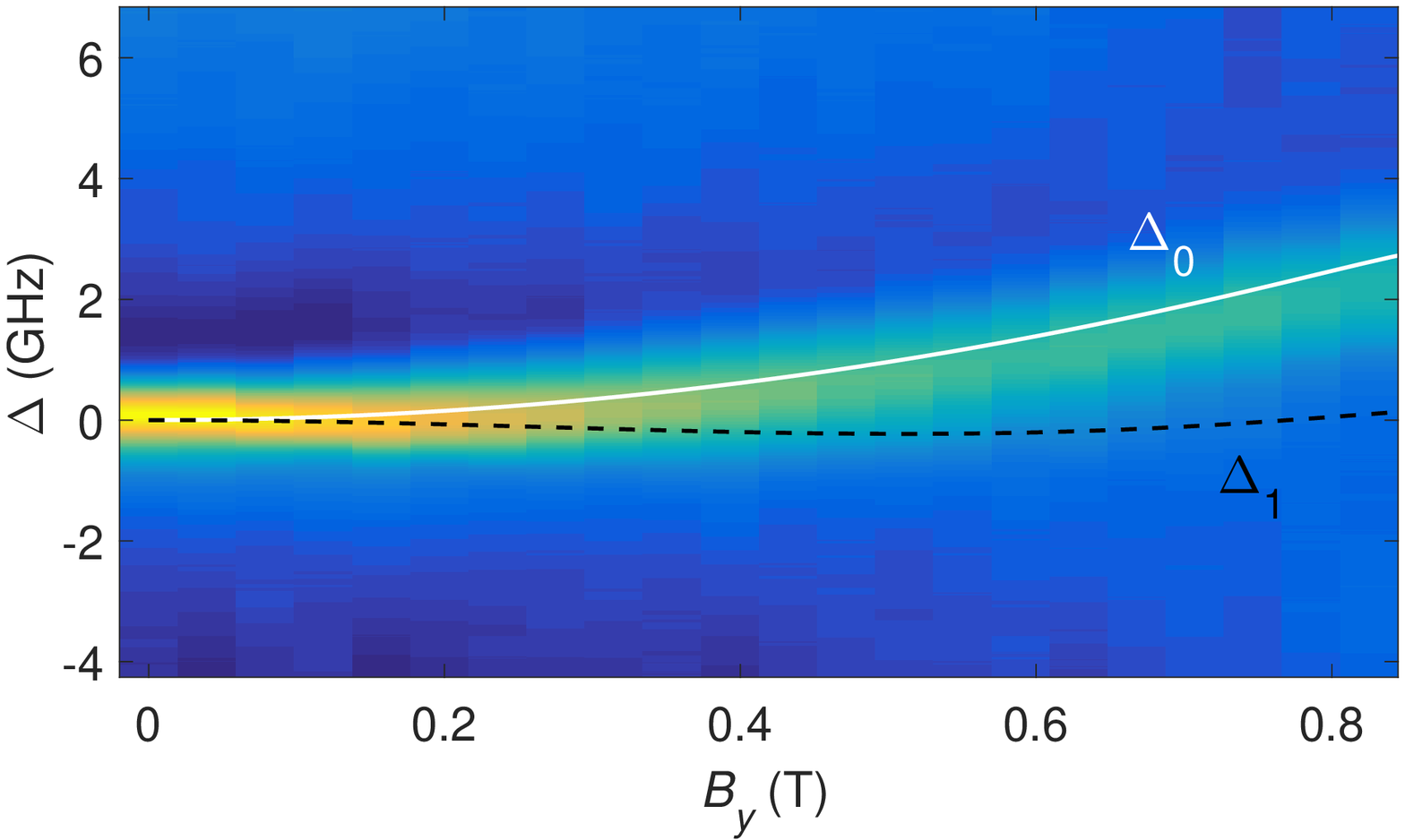}
	\caption{(color online) (top panel) Color map of the integrated bolometer signal versus the coupling magnetic field $B_y$ and tuning electric field $E_\bot$ for SE on liquid $^3$He at $T=0.33$~K and for electron density $n_s=5\times 10^6$~cm$^{-2}$. The perpendicular component of magnetic field is fixed at $B_z=0.584$~T. (bottom panel) The same data replotted versus the detuning (in GHz) from the center of the absorption line at $B_y=0$. For the sake of comparison, the theoretical results for the transition frequency shifts $\Delta_0$ and $\Delta_1$ (see Section~\ref{first}) obtained by the numerical diagonalization of the Hamiltonian \eqref{sec1:HtiltB} are shown by the white solid line and black dashed line, respectively.}
	\label{fig:6}
\end{figure}

To observe effect of the tilted magnetic field on the energy spectrum of SE, we recorded MW absorption measured by the above method either for a fixed value of $B_z$ and different values of $B_y$, or for a fixed value of $B_y$ and different values of $B_z$. The results are presented as 2D color maps of the integrated bolometer signal versus the varying component of $B$-field (horizontal axis) and $E_\bot$-field (vertical axis). The top panel of Fig.~\ref{fig:6} shows an example of such a plot illustrating the Lamb shift of the transition line with increasing coupling between the vertical motion and in-plane cyclotron motion. Here, the integrated bolometer signal taken for SE under MW excitation at $\omega/2\pi = 90$~GHz and at a fixed value of the perpendicular magnetic field $B_z=0.584$~T is plotted for different values of the coupling field $B_y$. At $B_y=0$ (no coupling), the $n=1\rightarrow 2$ transition occurs at $E_\bot\approx 23$~V/cm, which is in a reasonable agreement with the expected transition frequency $\omega_{21}/2\pi =(E_2 -E_1)/h = 90$~GHz for the Stark-shifted energy levels of electrons on liquid $^3$He~\cite{Issh2007}. With increasing $B_y$, the absorption line shifts towards lower values of $E_\bot$. This means that in order to have the same transition frequency $\omega_{21}$ between $n=1$ and $n=2$ Rydberg states, the energy levels must experience an additional shift in order to compensate for the reduction in the Stark shift due to $E_\bot$. The corresponding shift in the transition frequency is plotted in the bottom panel of Fig.~\ref{fig:6}. To obtain conversion between the shift in $E_\bot$ and the corresponding shift in $\omega_{21}$, we used the numerical value of $\kappa=0.74$~(GHz$\cdot$cm)/V for the slope of $\omega_{21}$ versus $E_\bot$ dependence, which was obtained in the experiment by measuring the absorption line for different values of the excitation frequency $\omega$. For the sake of comparison, the Lamb shift $\Delta_0$ and the light shift $\Delta_1$ due to the coupling of the Rydberg states to the in-plane motion, which were calculated as described in Section~\ref{first}, are plotted by the solid and dashed line, respectively. It is clear that the observed shift in the transition frequency corresponds to the Lamb shift $\Delta_0$, which implies that SE mostly occupy the lowest ($l=0$) Landau level. Indeed, at $B_z=0.584$~T the energy splitting between the ground $l=0$ state and the first excited $l=1$ state of the cyclotron motion is about $0.78$~K, which results in the thermal population of the ground state of about 92~$\%$. However, note that there is a barely visible splitting in the absorption line around $B_y\sim 0.5$~K. This splitting most likely appears due to transition of the small fraction of electrons occupying the first excited $l=1$ state. These electrons experience the transition frequency shift $\Delta_1$ of an opposite sign to $\Delta_0$, see dashed line in Fig.~\ref{fig:6}. At $B_y\sim 0.5$~T, the difference between $\Delta_0$ and $\Delta_1$ becomes comparable to the linewidth of the absorption line, which results in its splitting. We note that this splitting was more clearly observed in Ref.~\cite{Chep2019}, where significantly larger MW power was used in order to observe the transition line by a photo-assisted transport spectroscopy method.   

We note that broadening of the transition line increases with the increasing in-plane magnetic field $B_y$, as clearly seen in Fig.~\ref{fig:6}. This broadening is due to the many-electron effects and can be qualitatively understood using a simplified semi-classical argument. Due to the Coulomb interaction between electrons each electron experiences an instantaneous  fluctuating electric field $\textbf{E}_f(t)$, with the Gaussian distribution~\cite{Dykm1979}. In such a field the electron moves with the in-plane drift velocity $u=E_f/B_z$, which produces the Lorentz force $e u B_y$ acting on the electron in $z$-direction. The broadening of the transition line can be estimated as $eB_y \langle E_f \rangle/B_z$, where $\langle E_f \rangle \propto n_e^{3/4}$ is the rms Gaussian width of the fluctuating electric field~\cite{Dykm1997}. Thus,  the broadening is expected to increase with the in-plane magnetic field $B_y$. A detailed quantum theory of the many-electron broadening in tilted magnetic fields is beyond the scope of this work and will be reported elsewhere.        

\begin{figure}
\includegraphics[width=8.6cm]{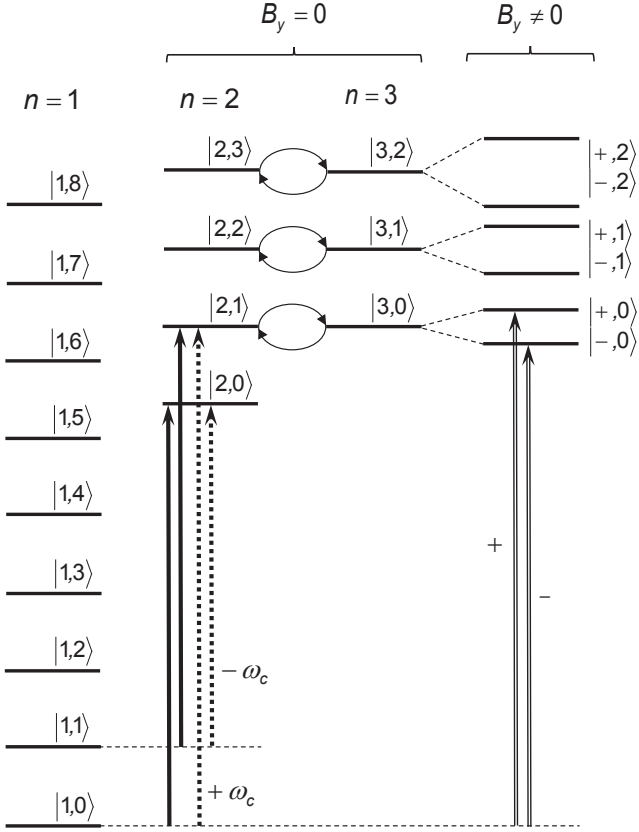}
\caption{(color online) Schematic energy level diagram for three lowest Rydberg state manifolds of an electron in a perpendicular magnetic field. Each manifold consists of the equidistant Landau levels $\hbar\omega_c (l+1/2)$, $l=0,1,..$ . For $B_y\neq 0$, the coupling leads to mixing between the states of different manifolds. Arrows show possible transitions between the dressed states, as described in the text. } 
\label{fig:7}       
\end{figure}

So far, we described MW-induced transitions of SE for which the quantum number $l$ is conserved. These transitions (for $l=0$ and 1) are showed by solid arrows in Fig.~\ref{fig:7}, which presents the schematic diagram of energy levels for three lowest Rydberg state manifolds of an electron in a perpendicular magnetic field. Without the in-plane component of magnetic field (no coupling between the vertical and in-plane motional states), resonant transitions between states which change $l$ are forbidden by the selection rules (we disregard the weak interaction of SE with ripplons which can lead to the second-order photon-assisted scattering processes). The coupling induced by a non-zero $B_y$ leads to mixing of different states, therefore a non-zero transition dipole moment for two arbitrary states of different manifolds. It is easy to see it by considering the situation far from the level crossing (the dispersive regime of CQED, see Section~\ref{first}) and apply first-order perturbation theory. As earlier, it is convenient to use the eigenstates $|n,l\rangle$ of the Hamiltonian \eqref{sec1:HperpB} as an unperturbed basis, therefore treat $H_1=H_I+m_e\omega_y^2z^2/2$ as the perturbation. To the lowest order in $B_y$, the admixed (dressed) state reads

\begin{equation}
|n,l\rangle_\textrm{ad} = |n,l\rangle + \frac{\hbar\omega_y}{\sqrt{2}l_B} \sum\limits_{n'} \frac{z_{nn'}\sqrt{l+\frac{1}{2}(1\pm 1)}}{E_{nn'}\mp \hbar\omega_c} |n',l\pm 1\rangle.
\label{sec2:dress}
\end{equation}  

\begin{figure}
\includegraphics[width=8.6cm]{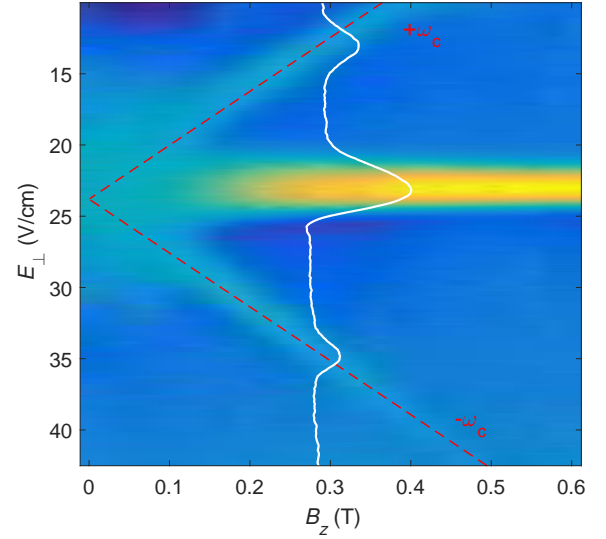}
\caption{(color online) MW absorption versus the perpendicular component of magnetic field $B_z$ and tuning electric field $E_\bot$ (scale reversed) taken for SE irradiated by MWs with frequency $\omega/2\pi=90$~GHz at $T=0.37$~K and for electron density $n_s=10^7$~cm$^{-2}$. The in-plane component of magnetic field is fixed at $B_y=0.2$~T. Two sideband transition branches are marked by $\pm\omega_c$. The red dashed lines plot $\kappa^{-1} \left( \omega_{21} \pm \omega_c\right)$, where $\kappa=0.74$~(GHz$\cdot$cm)/V is the conversion factor between the transition frequency $\omega_{21}$ and $E_\perp$. The white solid line shows a slice of the color map taken at $B_z=0.29$~T.} 
\label{fig:8}       
\end{figure}            

\noindent It is clear that for SE occupying states $|1,l\rangle$ of the first Rydberg state manifold, there exists nonvanishing probability for transitions to states $|n,l\pm 1\rangle_\textrm{ad}$ of higher ($n>1$) Rydberg state manifolds. Two such possible transitions are indicated in Fig.~\ref{fig:7} by dotted arrows and marked as $\pm \omega_c$. In an experiment, these transitions should result in two sideband absorption peaks with their transition frequencies separated by $\pm\omega_c$ from the absorption peak due to transitions which conserves the quantum number $l$ (solid arrows in Fig.~\ref{fig:7}). Figure~\ref{fig:8} shows a color map of the measured absorption versus the perpendicular component of magnetic field $B_z$ and tuning electric field $E_\bot$ taken for SE under MW excitation at the frequency $\omega/2\pi=90$~GHz and at a fixed value of the coupling field $B_y=0.2$~T. In addition to the strong absorption signal centered at $E_\perp^{(0)}\approx 23$~V, which corresponds to the transition $|1,l\rangle \rightarrow |2,l\rangle$, there are two sideband absorption peaks located on the opposite sides from the main peak. We note that such sideband peaks has been also seeing in the early spectroscopic experiments by Zipfel {\it et al.}~\cite{Zipf1976}. The dashed lines in Fig.~\ref{fig:8} represent dependance $\kappa^{-1} \left( \omega_{21} \pm \omega_c\right)$, which confirm that the frequencies of two sideband absorption peaks are separated by $\pm\omega_c$ from the main peak. Note that in addition to the sideband transitions which are accompanied by change of the quantum number $l$ by $\pm 1$, there should exist also much weaker higher-order sideband transitions which are accompanied by change of the quantum number $l$ by $\pm 2$, $\pm 3$, etc. 

We note that at sufficiently low values of $B_z \lesssim 0.2$~T the absorption line is smeared in the presence of non-zero $B_y$, as clearly seen in Fig.~\ref{fig:8}. This effect has been observed earlier by Zipfel {\it et al.} for SE in the in-plane magnetic field ($B_z=0$) and can be understood using the following simplified argument~\cite{Grim1976}. As an electron moves randomly along the surface of liquid helium with the thermal velocity $v_x$ in $x$-direction, it experiences the Lorentz force $e v_x B_y$ in $z$-direction due to the in-plane field $B_y$. This leads to the fluctuating effective electric field in $z$-direction with rms amplitude $\langle E_\bot \rangle =(k_B T/m_e)^{1/2}B_y$. This results in broadening of the transition line and strong reduction of the absorption near $B_z =0$, in agreement with our data shown in Fig.~\ref{fig:8}. Application of a sufficiently large perpendicular magnetic field strongly affects the electron in-plane motion, as discussed earlier. The full account of $B_z$-field on the transition line broadening, including the many-electron effects, will be given elsewhere. 

\begin{figure}[h]
	\includegraphics[width=8.6cm]{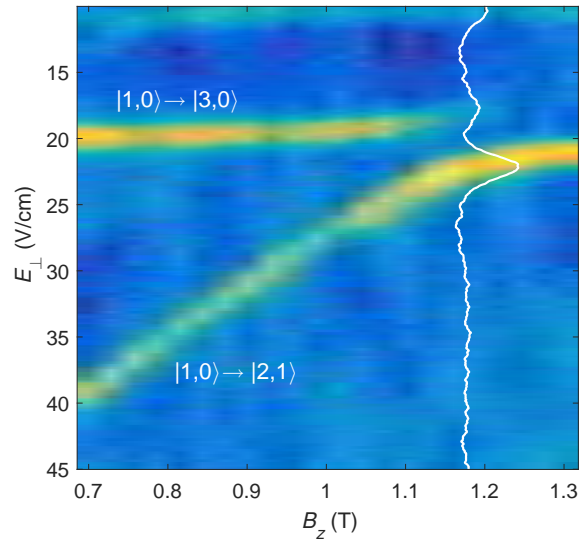}
	\caption{(color online) MW absorption versus the perpendicular component of magnetic field $B_z$ and tuning electric field $E_\bot$ (scale reversed) taken for SE irradiated by MWs with frequency $\omega/2\pi=120.5$~GHz at $T=0.35$~K and for electron density $n_s=10^7$~cm$^{-2}$. The in-plane component of magnetic field is fixed at $B_y=0.2$~T. The direct transition $|1,0\rangle \rightarrow |3,0\rangle$ and the sideband transition $|1,0\rangle \rightarrow |2,1\rangle$ are indicated for clarity. The white solid line shows a slice of the color map taken at the level crossing field $B_z=1.18$~T, illustrating the avoided crossing between two transition lines. An additional absorption peak at $E_\bot\approx 10$~V/cm is due to the resonant $|1,0\rangle\rightarrow |4,0\rangle$ transition.}
	\label{fig:9}
\end{figure}

\begin{figure}[h]
	\centering
	\includegraphics[width=0.85\linewidth]{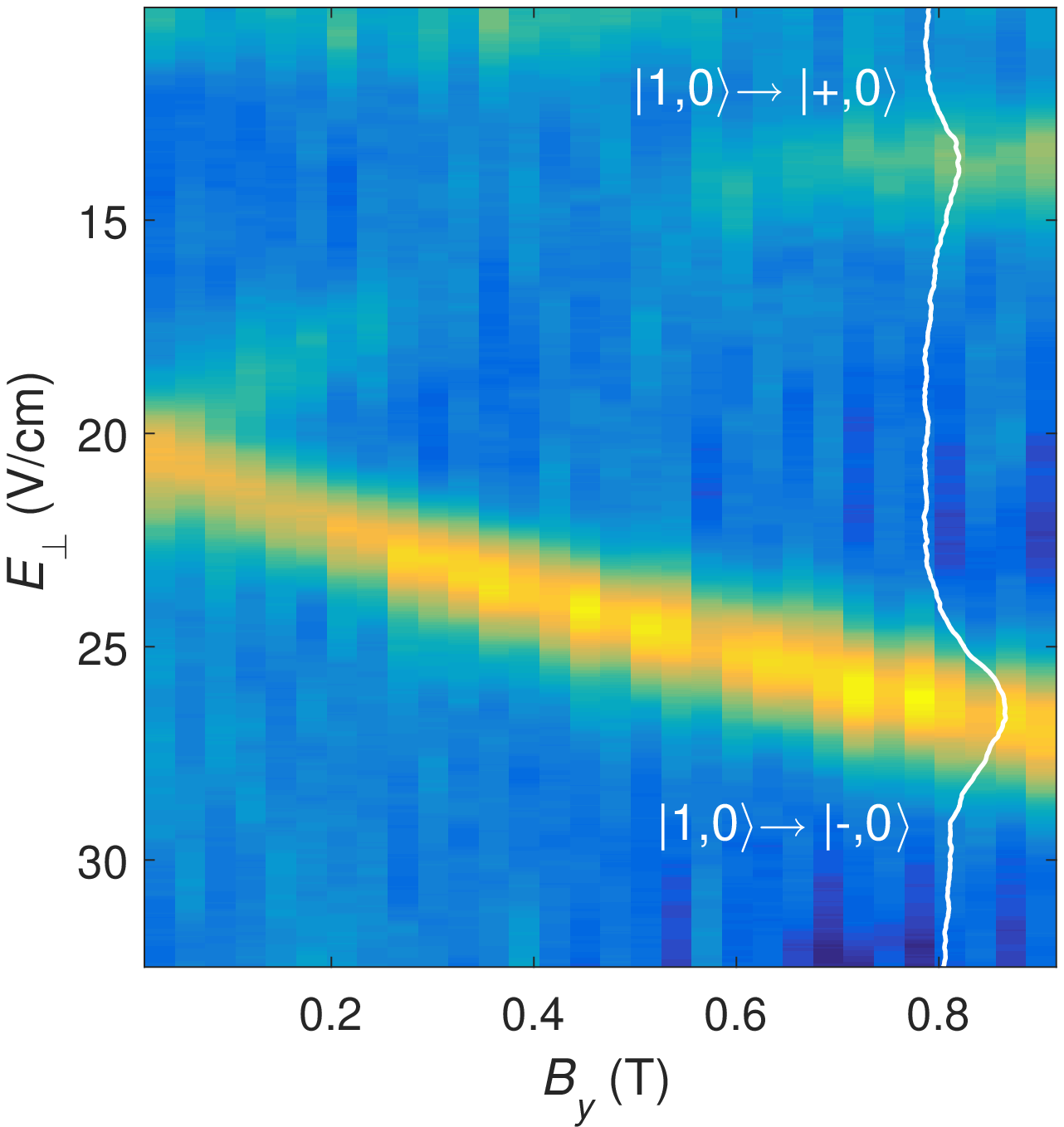}
	\includegraphics[width=0.84\linewidth]{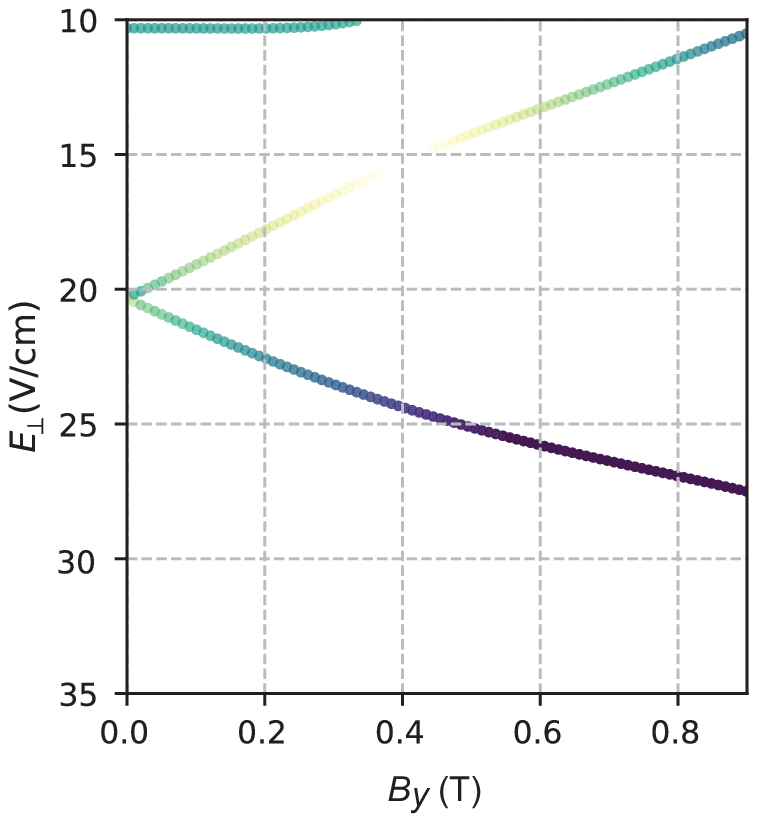}
	\caption{(color online) (top panel) MW absorption versus the coupling field $B_y$ and tuning field $E_\bot$ (scale reversed) taken for SE under the same conditions as in Fig.~\ref{fig:9}. The perpendicular component of magnetic field is fixed at $B_z=1.18$~T. Two branches corresponding to the transitions from the ground state $|1,0\rangle$ to hybridized states $|-,0\rangle$ and $|+,0\rangle$ are indicated for clarity. The white solid line shows a slice of the color map taken at $B_y=0.8$~T, illustrating the Autler-Townes absorption doublet. (bottom panel) The position of two transition lines of the doublet obtained from the diagonlization of the full Hamiltonian \eqref{sec1:HtiltB}. The calculated square of the dipole transition moment $z_{\pm}$ is given by the color tone for each line, as described in the text. The darker tone corresponds to the larger value of $|z_{\pm}|^2$.}
	\label{fig:10}
\end{figure}

To conclude this section, we consider the effect of coupling between the states $|n,l+1\rangle$ and $|n',l\rangle$ near their energy level crossing. As discussed in Section~\ref{first}, this corresponds to the resonant regime of coupling in CQED, the hallmark of which is an avoided level crossing for the dressed eigenstates. This effect is illustrated in Fig.~\ref{fig:7} for $n=2$ and $n'=3$ (see also Fig.~\ref{fig:3} in Section~\ref{first}). At $B_y=0$, the Landau levels of $n=2$ and $n=3$ Rydberg state manifolds are aligned, so each energy value is twice degenerate. At $B_y\neq 0$, the coupling  leads to a pair of hybridized states $|\pm,l\rangle$ (with corresponding energies $E_+\neq E_-$) for each $l=0,1,..$ . Thus, in a spectroscopic experiment one expects to observe the absorption doublets corresponding to the MW-induced transitions from the ground state $|1,0\rangle$ to the pair of states $|\pm,l\rangle$. An example of such a doublet is shown in Fig.~\ref{fig:7} by two double arrows marked as $\pm$. This is analogous to the Autler-Townes doublet observed in atomic and molecular systems, as well as in other quantum systems such as quantum dots, in a pump-probe experiment where a strong (pump) electromagnetic field induces repulsion between two dressed atomic levels (the Autler-Townes splitting), while another weaker field probes transitions between a third levels and the dressed states~\cite{Cohen_book,Wrig2007,Sham2007}. The repulsion between dressed states and the Autler-Townes doublet are also observed in our experiment. Figure~\ref{fig:9} shows an example of the measured absorption versus the perpendicular component of magnetic field $B_z$ and tuning electric field $E_\bot$ taken for SE under MW excitation at frequency $\omega/2\pi=120.5$~GHz and at a fixed value of the coupling field $B_y=0.2$~T. The transition $|1,0\rangle\rightarrow |3,0\rangle$ results in an absorption peak centered at $E_\bot\approx 20$~V/cm. Its position is almost independent of $B_z$. The position of the sideband transition $|1,0\rangle\rightarrow |2,1\rangle$  changes linearly with $B_z$. The corresponding transition frequencies cross at $B_z=1.18$~T, near which there is an avoided crossing between absorption peaks. To illustrate it further, Fig.~\ref{fig:10} (top panel) shows a color map of the measured absorption taken at the level crossing field $B_z=1.18$~T for different values of the coupling field $B_y$. As $B_y$ increases from zero, the absorption peak centered at $E_\bot=20$~V/cm splits into two peaks. The top and bottom branches of this Autler-Townes doublet correspond to the transitions of SE from the ground state $|1,0\rangle$ to the hybridized states $|+,0\rangle$ and $|-,0\rangle$, respectively (see Fig.~\ref{fig:7}). As discussed in Section~\ref{first}, the energy splitting between two lines of the doublet is approximately given by $2g_{23}=\sqrt{2\hbar\omega_c}\omega_yz_{23}$, which increases linearly with $B_y$, although some deviations from this analytical expression come from the higher order mixing between eigenstates. For the sake of comparison, the position of transition lines obtained from the energy spectrum calculated by diagonalization of the Hamiltonian \eqref{sec1:HtiltB} are shown in the bottom panel of Fig.~\ref{fig:10}.  

An interesting feature of the data presented in Figs.~\ref{fig:9} and \ref{fig:10} is that two lines of the doublet show different values for the absorption rate. In particular, the upper branch of the doublet shows lower absorption, which also varies with the coupling field $B_y$. Surprisingly, in a certain window of $B_y$ values around 0.4~T the absorption disappears, that is the system becomes transparent for the microwave radiation. This effect appears due to quantum interference of the probability amplitudes of the dipole-allowed transitions. According to Fermi's golden rule, the probability of transitions from the ground state $|1,0\rangle$ to the excited dressed states $|\pm,0\rangle$ are proportional to $|z_{\pm}|^2$, respectively, with the transition dipole moments $z_{\pm}=\langle \pm,0|z|1,0\rangle$. As discussed in Section~\ref{first}, in the two-level approximation the dressed states are $|\pm,0 \rangle =2^{-1/2} \left( |2,1\rangle \pm |3,0\rangle \right)$. Beyond the two-level approximation, the states $|2,1\rangle$ and $|3,0\rangle$ are also admixed with other states. In particular, according to \eqref{sec2:dress} up to the leading order in $\hbar\omega_y/R_e$ the admixed states read

\begin{eqnarray}
&& |2,1\rangle_\textrm{ad} \approx |2,1\rangle + \frac{z_{22}}{\sqrt{2}l_B} \left( \frac{B_y}{B_z} \right) |2,0\rangle - \frac{z_{22}}{l_B} \left( \frac{B_y}{B_z}\right) |2,3\rangle, \nonumber \\
&& |3,0\rangle_\textrm{ad} \approx |3,0\rangle - \frac{z_{33}}{l_B} \left(\frac{B_y}{B_z}\right) |3,1\rangle.
\label{sec3:interfer}
\end{eqnarray}

\noindent Thus, the transition dipole moments to the leading order are given by

\begin{equation}
z_{\pm}=\frac{z_{22}z_{21}}{\sqrt{2}l_B} \left( \frac{B_y}{B_z} \right)  \pm z_{31}.
\label{sec3:zpm}
\end{equation}

\noindent It is clear that there is either a cancellation or an enhancement of the above dipole moments, therefore the corresponding transition probabilities,  depending on the relative sign between the matrix elements $z_{21}$ and $z_{31}$. Using numerical estimates $|z_{31}/z_{21}|\approx 0.5$ and $z_{22}\approx 3.91 r_B$ valid for the conditions of our experiment, the complete cancellation in \eqref{sec3:zpm} occurs at $B_y=0.49$~T, which agrees reasonably well with Fig.\ref{fig:10}. The numerically calculated squares of the transition moments $z_{\pm}$ for two branches are plotted in the bottom panel of Fig.~\ref{fig:10} as the color tone for each line of the doublet. The darker tone corresponds to the larger value of $|z_{\pm}|^2$. Comparison with the top panel of Fig.~\ref{fig:10} shows that the variation of the measured absorption rate for two branches is in a very good agreement with the corresponding variation of the transition probability. 

We note that this effect is reminiscent of the electromagnetically induced transparency (EIT) observed in the two-photon spectroscopy of three-level systems~\cite{Imam2005,Mara1998}. In such experiments, two energy levels are coupled by a resonant electromagnetic (control) field, while another weak (probe) field is tuned for the transition between one of this level and a third level. Similar to our case, significant suppression of the probe field absorption by the system in such experiments can be explained by the destructive interference of the dipole-allowed transitions between the third level and the dressed states of the two levels coupled by the control field~\cite{Mara1998}.             



\section{Discussion and conclusion}\label{third}

We have shown that our straightforward calculations based on the numerical diagonalization of the full Hamiltonian \eqref{sec1:HtiltB} are able to reproduce all features of the measured spectroscopic properties of SE very well. It is important to emphasize that the JCM Hamiltonian \eqref{sec1:HtiltB} is completely controlled only by the values of dc electric (via the Stark effect) and magnetic (via the Landau quantization and tunable coupling between the motional states) fields used in the experiment, therefore the calculations contain no adjustable parameters. This allows us to conclude that the considered system of SE in a tilted magnetic field can serve as a convenient and robust platform complementary to atomic and molecular systems, such as trapped ions and Rydberg atoms in cavities, to realize JCM and study related quantum phenomena. 

\begin{figure}[h]
\includegraphics[width=7cm]{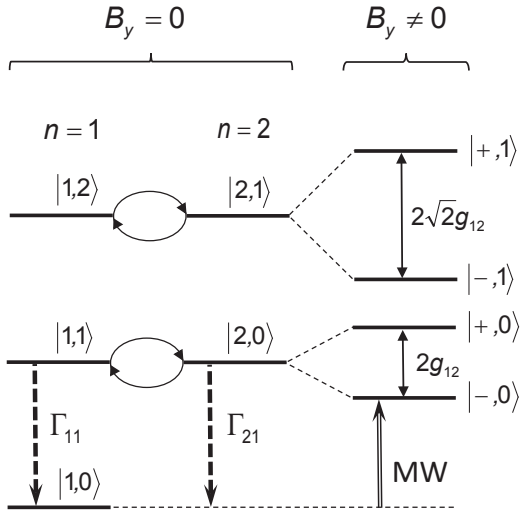}
\caption{(color online) Schematic energy level diagram at the energy crossing for states $|1,1\rangle$ and $|2,0\rangle$. $\Gamma_{11}$ and $\Gamma_{21}$ represent decay rates of these states due to inelastic two-ripplon scattering, as discussed in the text. The double arrow indicates the transition between the ground state $|1,0\rangle$ and the hybridized state $|-,0\rangle$, which can be induced by the microwave electric field polarized parallel to the surface of liquid (see further discussion in the text).}
\label{fig:12}
\end{figure} 

As a particular example, let us consider the regime of resonant coupling near the energy crossing for states $|1,1\rangle$ and $|2,0\rangle$, see Fig.~\ref{fig:12}. For SE on liquid $^3$He in a perpendicular electric field $E_\bot=15$~V/cm such a level crossing can be realized with an applied perpendicular magnetic field $B_z\approx 2.82$~T, see Fig.~\ref{fig:2}. JCM predicts that if an electron is initially prepared in the state $|2,0\rangle$ and is suddenly brought to the situation where this state crosses levels with the state $|1,1\rangle$, for example by applying a Stark pulse of the electric field $E_\bot$, the state of electron undergoes the coherent (Rabi) oscillations between $|2,0\rangle$ and $|1,1\rangle$. In particular, the system oscillates between zero and one quantum of the in-plane cyclotron motion. This is an analog of the vacuum Rabi oscillations in CQED. The Rabi frequency of oscillations is given by $g_{12}/h$. According to Eq.~\eqref{sec1:g}, it is the order of 10~GHz at the coupling field $B_y\approx 1.5$~T. The regime of strong coupling, where the system undergoes many cycles of such reversible oscillations between two states, can be reached providing that the relaxation and dephasing rates for these states are much smaller than the Rabi frequency. Thus, it is important to provide an estimation for these rates. 

At sufficiently low temperatures (below 1~K), the dissipative processes in SE are mostly due to their coupling to the capillary waves excited on the surface of liquid helium. These waves introduce distortion of the surface of the liquid, therefore modifying the electron potential energy \eqref{sec1:V}. In particular, the first term in \eqref{sec1:V} becomes $V_0\Theta (\xi-z)$, where $\xi(\textbf{r})$ is the surface displacement which is a function of the electron position vector $\textbf{r}$ in the $xy$-plane. The standard quantization procedure allows us to represent $\xi$ as

\begin{equation}
\xi=\sum\limits_{\textbf{q}} Q_q e^{i\textbf{q}\textbf{r}} (b_{\textbf{q}} + b_{-\textbf{q}}^\dagger),
\end{equation}

\noindent where $Q_q=\sqrt{\hbar q/(2S\rho \omega_q)}$, $S$ is the surface area, and Bose operators $b_{\textbf{q}}$ and $b_{\textbf{q}}^\dagger$ describe ripplons with the usual capillary wave dispersion $\omega_q=\sqrt{\alpha q^3/\rho}$ ( here $\alpha$ is the surface tension of liquid and $\rho$ is the liquid density). The second term in \eqref{sec1:V}, which corresponds to the polarization attraction of an electron to helium atoms comprising the liquid, has an integral form which depends on $\xi$. Since the mean displacement of surface $\langle\xi\rangle\lesssim 1$~$\AA$ is much smaller than the mean distance $\langle z \rangle$ between an electron and the surface, it is conventional to expand the potential energy of electron in $\xi$ and apply the perturbation theory~\cite{Monarkha_book}. The term linear in $\xi$ is responsible for the one-ripplon scattering processes, which are almost elastic due to the softness of ripplon modes. The dissipative processes in SE are mostly due to two-ripplon scattering, in particular the spontaneous emission of couples of short-wavelength ripplons. These processes are usually accounted for by the term quadratic in $\xi$ in the first order of the perturbation theory.~\cite{Mona1978} The strongest contribution to this process comes from the presence of the large repulsive barrier $V_0$, with a corresponding term in the electron-ripplon interaction Hamiltonian given by

\begin{equation}
V_{\textrm{int}}^{(2)} (z,\textbf{r})= \frac{1}{2} \frac{\partial V_0\Theta (\xi-z)}{\partial z} \Bigg|_{\xi=0} \xi(\textbf{r})^2.
\end{equation} 

\noindent According to Fermi's golden rule, the rate of decay from a state $|n,l\rangle$ into a lower-energy state $|n',l'\rangle$ due to the two-ripplon emission is then given by

\begin{eqnarray}
&& \Gamma_{nn'} = \frac{\pi}{\hbar} \left( \frac{\partial V_0\Theta (\xi-z)}{\partial z} \Bigg|_{\xi=0} \right)^2_{n,n'}\sum\limits_{\textbf{q},\textbf{q}'} Q_q^2 Q_{q'}^2 (N_q+1)(N_{q'}+1)  \nonumber \\
&& \times |(e^{i(\textbf{q}'-\textbf{q})\textbf{r}})_{ll'}|^2\delta (E_{n',l'}-E_{n,l}+\hbar\omega_q + \hbar\omega_{q'}), 
\label{sec3:G}
\end{eqnarray}

\noindent where $N_q$ is the average ripplon occupation number. An important point regarding the two-ripplon emission process is that, because of the softness of ripplons modes, the conservation of energy and momentum requires the total momentum of two ripplons to be small, that is $\textbf{q}\approx -\textbf{q}'$~\cite{Mona1978}. This introduces significant simplifications for the estimation of the above expression for the decay rate. In addition, for practical purposes it is convenient to represent the matrix element of interaction appearing in \eqref{sec3:G} as

\begin{eqnarray}
&& \left( \frac{\partial V_0\Theta (\xi-z)}{\partial z} \Bigg|_{\xi=0} \right)_{n,n'} = \sqrt{\frac{2V_0\hbar^2}{m_e}}\psi'_n(0)\psi'_{n'}(0) \nonumber \\
&& = \frac{\sqrt{8m_eV_0}}{\hbar} \sqrt{\left(\frac{\partial \upsilon}{\partial z}\right)_{nn}\left(\frac{\partial \upsilon}{\partial z}\right)_{n'n'}}.
\label{sec3:matr}
\end{eqnarray} 

\noindent Here, $\psi_n(z)$ is the wavefunction for the motion perpendicular to the surface and $\upsilon(z)=-\Lambda/z + e E_{\bot} z$. Note that for the numerical estimation of the matrix elements in \eqref{sec3:matr} it is sufficient to use the wavefunctions corresponding to the rigid-wall repulsive barrier.

Using the above simplifications, it is straightforward to perform summations in \eqref{sec3:G} and estimate the decay rate. In particular, the rates $\Gamma_{21}$ and $\Gamma_{11}$ for the decay of the excited states $|2,0\rangle$ and $|1,1\rangle$ (see Fig.~\ref{fig:12}), respectively, can be represented as~\cite{Yuri2007,Mona1978}  

\begin{eqnarray}
&& \Gamma_{21} = \frac{m_e V_0}{4\pi l_B^2\rho^2\hbar^2} \left(\frac{\partial \upsilon}{\partial z}\right)_{11}\left(\frac{\partial \upsilon}{\partial z}\right)_{22} \frac{\tilde{q}^3}{\omega_{\tilde{q}}^2 |\partial\omega_{\tilde{q}}/\partial \tilde{q}|}, \\
\label{sec3:matr1}
&& \Gamma_{11} = \frac{m_e V_0}{4\pi l_B^2\rho^2\hbar^2} \left(\frac{\partial \upsilon}{\partial z}\right)^2_{11} \frac{\tilde{q}^3}{\omega_{\tilde{q}}^2 |\partial\omega_{\tilde{q}}/\partial \tilde{q}|}.
\label{sec3:matr2}
\end{eqnarray}       

\noindent Here, $\tilde{q}$ is the ripplon wave number which satisfies the energy conservation relation $2\hbar\omega_{\tilde{q}}=E_2-E_1$. For SE on liquid $^3$He in the perpendicular electric field $15$~V/cm we have $\tilde{q}\approx 3\times 10^7$~cm$^{-1}$. Then, numerical evaluation of the above expressions give $\Gamma_{21}\approx 6\times 10^5$~s$^{-1}$ and $\Gamma_{11}\approx 1.4\times 10^6$~s$^{-1}$, which corresponds to the lifetime of an excited state of the order 1~$\mu$s. We note that this estimate agrees very well with the experimentally observed relaxation time of the excited Rydberg states of SE in zero magnetic field~\cite{Kawa2020}.  

The first-order elastic one-ripplon processes lead to the stronger scattering of SE between the degenerate states of each energy level. The rate of this process can be roughly estimated using the self-consistent Born approximation (SCBA) and the known rate of the elastic one-ripplon scattering $\nu_0$ in zero magnetic field. According to SCBA, the elastic scattering rate $\nu_B$ in non-zero magnetic field is enhanced by a factor of $\hbar\omega_c/\Delta_c$, where $\Delta_c$ is the collision-broadened width of the Landau levels. This leads to a simple relation $\nu_B=\sqrt{2\omega_c\nu_0/\pi}$~\cite{Ando1974}. The typical one-ripplon scattering rate at $T=0.1$~K is about $10^6$~s$^{-1}$, which gives $\nu_B\approx 5\times 10^8$~s$^{-1}$. This process will lead to the dephasing. For a many-electron system, the electron-electron interaction can lead to significant broadening of the Landau levels, therefore to reduction of the above elastic scattering rate $\nu_B$~\cite{Mona2012}. 

Thus, a single surface-bound electron on liquid helium in a tilted magnetic field realizes a quantum system with on-site JCM-type interaction where the coherent part of evolution can dominates over the dissipative processes. Another remarkable fact regarding electrons on helium is that at sufficiently low temperatures $T\leq 1$~K and moderate electron densities $n_s \leq 10^9$~cm$^{-2}$, SE crystallize into a triangular Wigner lattice~\cite{Monarkha_book}. In addition, the Coulomb interaction introduces an effective interaction between electrons associated with their vertical quantized motion. To the lowest order in $r_B/a$, where $a\approx n_s^{-1/2}$ is the inter-electron distance, this interaction is described by~\cite{Dykm2003,Kons2012} 

\begin{eqnarray}
&& H_{ee} = \frac{e^2 n_s^{3/2}}{16\pi\epsilon_0} \sum\limits_{j\neq j'} (z_j-z_{j'})^2  \\
&& \approx \frac{e^2 n_s^{3/2}}{16\pi\epsilon_0}  \sum\limits_{j\neq j'} \left[ (z_{22}-z_{11})^2 \sigma^{j}_z \sigma^{j'}_z + 4|z_{12}|^2 ( \sigma^{j}_+\sigma^{j'}_- + \textrm{h.c.} ) \right], \nonumber
\label{sec3:Hee}
\end{eqnarray} 

\noindent where $j$ and $j'$ are the lattice sites and the effective spin operators for each site are defined as $\sigma^{j}_z = |2\rangle_{jj}\langle 2 | - |1\rangle_{jj}\langle 1 |$ and $\sigma^{j}_+ = (\sigma^{j}_-)^\dagger = |2\rangle_{jj}\langle 1|$ (note that we neglected excitation for the $n>2$ Rydberg states). The two terms in the above Hamiltonian describe the static dipolar interaction and the excitation hopping between the lattice sites. Thus, SE system on liquid helium subject to a tilted magnetic field realizes a quantum many-body system with on-site interaction of JCM-type and strong interaction between the sites. This is an interesting complement to the Jaynes-Cummings lattice models, which has recently attracted attention due to the possibility to study quantum phase transitions in strongly-correlated systems~\cite{Plen2008}. The system described here, where the on-site interaction can be easily adjusted by the value of the applied magnetic field, while the inter-site interaction can be readily varied by changing the density of SE, presents a promising flexible platform to study such many-body models. In addition to the conventional absorption measurements described here, the image-charge detection method can provide a convenient way to measure the quantum state of SE~\cite{Kawa2019}. As was shown, this method potentially can be scaled to detect excitation of a single lattice site. 

Another interesting point about SE in a tilted magnetic field is that the coupling introduces a strong nonlinearity in the harmonic spectrum of the in-plane cyclotron motion, in particular when the Landau levels of two Rydberg manifolds align. Under this condition, the energy levels of the hybridized in-plane motion become strongly non-equidistant, and the resonant transition between the ground state $|1,0\rangle$ and one of the  hybridized state, e.g. $|-,0\rangle$ (see Fig.~\ref{fig:12}), can be excited by the MW radiation with electric field polarized parallel to the surface. This presents an opportunity for anther interesting CQED-type experiment with a many-electron ensemble on liquid helium coupled to a single-mode cavity resonator. Recently, the strong coupling of such an ensemble to a high-quality single-mode Fabry-Perot resonator has been demonstrated~\cite{Chen2018}. In this experiment, the resonator geometry favors the polarization of the microwave electric field being parallel to the surface, therefore the cyclotron motion of SE in a perpendicular magnetic field was excited. However, the linearity of such a coupled system precludes to observe differences between purely classical behaviour and any predictions based on quantum mechanics~\cite{Chen2018}. By introducing the parallel component of magnetic field, the strong coupling between an ensemble of the highly nonlinear two-level systems and a single mode of a cavity field can be readily realized and studied.

In conclusion, we study the motional quantum states of the surface-bound electrons on liquid helium subjected to a tilted magnetic field and show that they are described by the JCM Hamiltonian. The predictions of theory regarding spectroscopic properties of such a system show complete agreement with the experimental results, without using any adjustable parameters. The system shows many similarities with the quantum systems interacting with light, in particular a number of phenomena related to the ac Stark effect in atomic and molecular systems. Also, we predict that for moderately low temperatures and values of the coupling magnetic field the coherent evolution of coupled states dominates over the dissipative processes in the system. Thus, this system potentially presents a new robust and flexible platform for quantum experiments. Interestingly, our work introduces a pure condensed-matter system of electrons on helium into the context of atomic and optical physics, which might provide an opportunity to bridge different fields, for example many-body physics and quantum optics.    

{\bf Acknowledgements} The work was supported by an internal grant from Okinawa Institute of Science and Technology (OIST) Graduate University.



\end{document}